\documentclass[12pt,notoc]{JHEP3}

\usepackage{amsmath,amssymb,euscript,array,cite}

\input{epsf}
\newcommand{\startappendix}{
\setcounter{section}{0}
\renewcommand{\thesection}{\Alph{section}}}
\newcommand{\Appendix}[1]{
\refstepcounter{section}
\begin{flushleft}
{\large\bf Appendix \thesection: #1}
\end{flushleft}}
\usepackage{epsfig}

\def\N{{\cal N}}

\def\Dbarslash{\,\,{\raise.15ex\hbox{/}\mkern-12mu {\bar D}}}
\def\Dslash{\,\,{\raise.15ex\hbox{/}\mkern-12mu D}}
\def\delslash{\,\,{\raise.15ex\hbox{/}\mkern-9mu \partial}}
\def\delbarslash{\,\,{\raise.15ex\hbox{/}\mkern-9mu {\bar\partial}}}

\newcommand{\MAT}[1]{\begin{pmatrix} #1\end{pmatrix}}
\newcommand{\EQ}[1]{\begin{equation} #1 \end{equation}}

\newcommand{\SP}[1]{\begin{equation}\begin{split} #1
\end{split}\end{equation}}

\title{Double Scaling Limits in Gauge Theories and Matrix Models}
\author{Gaetano Bertoldi${}^1$, Timothy J. Hollowood${}^1$ and J. Luis Miramontes${}^2$\\
${}^1$ Department of Physics,\\ University of Wales Swansea,\\
Swansea, SA2 8PP, UK.\\
E-mail: {\tt g.bertoldi, t.hollowood@swan.ac.uk} \\
${}^2$ Departamento de F\'isica de Part\'iculas,\\ and Instituto Gallego de F\'isica de Altas Energ\'ias (IGFAE),\\
Facultad de F\'isica, Universidad de Santiago de Compostela \\
15782 Santiago de Compostela, Spain\\
E-mail: {\tt miramont@usc.es}  } 
\preprint{SWAT/06/459}
\abstract{We show that $\N=1$ gauge theories with an adjoint chiral
  multiplet admit a wide class of large-$N$
  double-scaling limits where $N$ is taken to infinity in a way
  coordinated with a tuning of the bare superpotential. The tuning is
  such that the theory is near an Argyres-Douglas-type singularity where a
  set of non-local dibaryons becomes massless in conjunction with a set
  of confining strings becoming tensionless. The
  doubly-scaled theory consists of two decoupled sectors, one whose
  spectrum and interactions follow the usual large-$N$ scaling whilst
  the other has light states of fixed mass in the large-$N$ limit
  which subvert the usual large-$N$ scaling and lead to an
  interacting theory in the limit. $F$-term properties of this
  interacting sector
  can be calculated using a Dijkgraaf-Vafa matrix model and in this context the
  double-scaling limit is precisely the kind investigated in the
  ``old matrix model'' to describe two-dimensional gravity
  coupled to $c<1$ conformal field theories. In particular, 
  the old matrix model double-scaling limit describes a sector of a 
  gauge theory with a mass gap and light meson-like composite states, 
  the approximate Goldstone boson of superconformal invariance, 
  with a mass which is fixed in the double-scaling limit. 
  Consequently, the gravitational $F$-terms in these cases satisfy the 
string equation of the KdV hierarchy.}


\begin{document}

\section{Introduction}

't Hooft argued many years ago \cite{TH1} that the perturbative expansion
of theories involving matrices has an interesting interpretation
when $N$, the size of the matrices, becomes large. The idea is to
first re-write the perturbative expansion in $g_s$, the 
coupling, and $N$, in terms of  
$S=g_sN$, the ``'t Hooft Coupling'', and $g_s$. The Feynman graphs
of the theory can be sorted on the basis of the 
topology of an imaginary two-dimensional surface on which the
diagrams can be drawn. In terms of $(S,g_s)$, the explicit 
$g_s$ dependence of a graph is determined solely by the topology of the 
imaginary
surface: if $g$ is the genus then the dependence is $g_s^{2g-2}$. At a given
genus, there is then a perturbative series in the 't Hooft coupling $S$.

An example of such matrix theories are Yang-Mills theories.
These ideas led 't~Hooft to suggest that Yang-Mills theories may have
an alternative description in which the imaginary two-dimensional
surfaces become physically realized as the world-sheets of a string
theory. In this picture, $g_s=g_\text{YM}^2$ 
becomes the {\it string coupling\/} weighting
each handle in string world-sheet perturbation theory, so that
the string theory is weakly coupled when $N$ is large. Quite how gauge
theories are equivalent to a theory of strings has taken many years to
understand in a concrete way.

The topological expansion of matrix field theories is quite general and
applies, for example, to simple matrix theories (zero-dimensional
field theories), where there
is no spacetime. The fact that such simple theories can give rise to
two-dimensional surfaces has been an idea of great import. In the ``old
matrix model'' epoch the idea was to use matrix models to
describe models of two-dimensional gravity coupled to various matter
systems. Here, once again the surfaces arise in an auxiliary way from
the expansion of the matrix model in the string coupling. 
In these theories the way
that the Feynman graphs give rise to a two-dimensional surface is
completely transparent. The 't Hooft expansion of, say, the free
energy has the form
\EQ{
F=\sum_{g=0}^\infty F_g(\xi)g_s^{2g-2}\ .
\label{deff}
}
Here, $\xi$ label a general set of parameters in the theory (suitably
scaled in $N$), including
the 't Hooft coupling. 
The surfaces arise when one takes the large-$N$ limit with fixed 't
Hooft coupling, as before, but,
in addition, one takes a parameter $\xi$ 
to a critical value. At the critical point
$\xi=\xi_0$, the contributions from a given genus $F_g(\xi)$ diverge in
a particular way:
\EQ{
F_g(\xi)\thicksim |\xi-\xi_0|^{\eta(2-2g)}\ ,
\label{fdsl}
}
where $\eta$ is a critical exponent
which characterizes the universality of the critical point. Intuitively, near
 the critical point the Feynman
graphs of a given genus become very dense and describe a surface. The
coordinated, or double-scaling limit, involves taking \cite{oldMMdsl}
\EQ{
N\to\infty\ ,\qquad \xi\to \xi_0\ ,\qquad \Delta=
N|\xi-\xi_0|^\eta=\text{fixed}\  .
\label{dssl}
}
In this limit there is an effective string coupling 
\EQ{
(g_s)_\text{eff}\thicksim \Delta^{-1}
\label{effg}
}
which weights the contributions from different genera.

What makes the old matrix model story so compelling is the ease at
which the two-dimensional surfaces arise. This is not so clear in 
Yang-Mills theories where the surfaces are thought to be associated to
a dual string description. Realizing this picture explicitly was
brilliantly achieved by Maldacena in the case of maximally
supersymmetric $\N=4$ gauge theories \cite{AdS}. In this case the dual string
theory is a ten-dimensional Type IIB superstring on the space
$AdS_5\times S^5$ with background Ramond-Ramond (RR) flux. Unfortunately
such string theories have proved very difficult to solve because of the
presence of the RR flux. 
It would be very useful to have a situation where
the string theory appeared in a very simple way as in the old matrix
model. What we have in mind is a realization of a double-scaling
limit in a Yang-Mills theory where $N$ is taken to infinity as 
certain couplings in a bare potential are taken
to critical values in such a way that the string dual naturally
arises. Double-scaling limits of this kind 
have been recently studied in \cite{BD} for a class of
$\N=1$ gauge theories in a  {\it partially confining\/} phase. 
There it was proposed that they admit a dual description 
in terms of a non-critical string theory of the type
first introduced in \cite{Kutasov:1990ua}. The worldsheet theory
corresponding to this string background is
exactly solvable and there are no RR fluxes. 
The gauge theories involve
a gauge group $G=U(N)$ which is confined down to an abelian group.
The theories have an $\N=1$ vector multiplet and an
adjoint-valued chiral multiplet $\Phi$ (precisely the fields of an
$\N=2$ theory).\footnote{The 
$\beta$-deformation of ${\cal N}=4$ Super Yang-Mills 
also has a partially confining phase that has been studied in 
\cite{nd1,nd2,DH}. A double-scaling limit and
the relative non-critical string dual have also been proposed in \cite{BD}.}
There is an explicit bare superpotential
\EQ{
W_\text{bare}(\Phi)=\text{Tr}\,W(\Phi)\ ,\qquad
W(x)=N\sum_{i=1}^{\ell+1}\frac{g_i}i x^i\ .
\label{bsup}
}
The prefactor ensures conventional large-$N$ scaling.
Classically, each of the eigenvalues of the scalar component of $\Phi$
can be one of the $\ell$ critical points of $W(x)$,
\EQ{
W'(x)=N\varepsilon\prod_{i=1}^\ell(x-a_i)\ ,\qquad\varepsilon\equiv
g_{\ell+1}\ .
}
We define the multiplicity of eigenvalues at the critical point $a_i$
as $N_i\geq0$.
At least in an appropriate limit, this leads to a classical Higgs effect
$U(N)\to\prod_{i=1}^\ell U(N_i)$, where $\sum_{i=1}^\ell N_i=N$. Each $U(N_i)$
will then subsequently confine to leave a $U(1)$ factor in the IR. We
will denote by $s$ the number of $N_i>0$, so that the IR gauge
group is $U(1)^{s}$. Of course, such
a picture of a classical Higgs effect plus confinement is not expected
to be valid for all the values of the parameters but is a useful limit
to have in mind. In particular, vacua with the same rank of IR group, 
but with different ``filling fractions'' $N_i/N$, are known, under
certain circumstances, to be continuously connected in the space of
superpotentials \cite{Friedmann:2002ct}\cite{CSW}.

The spectrum of these partially confining theories is expected to be rather
complicated and since there is only $\N=1$ supersymmetry and
no notion of a BPS state, we have
little analytic control over particle masses.
However, we expect that for a generic vacuum the
conventional large-$N$ picture applies (see for example \cite{man}): 
as well as the massless abelian sector, there
will be a tower of glueballs and "dimesons", 
formed by one "diquark" $Q_{rs}$ transforming
in the $(N_r,N_s)$ representation of $U(N_r) \times U(N_s)$ and the
corresponding  anti-diquark $\bar{Q}_{rs}$, whose masses
are $N$ independent. There will also be dibaryons, whose masses scale like $N$. 
The interactions of the glueballs and mesons are suppressed by
$1/N$,\footnote{The 
interactions between mesons are suppressed
by integer powers of $1/N$ instead of the usual $1/\sqrt{N}$ because
these states are formed by one "diquark" $Q_{rs}$ transforming
in the $(N_r,N_s)$ representation of $U(N_r) \times U(N_s)$ and the
corresponding  anti-diquark $\bar{Q}_{rs}$.  Therefore the three-point
coupling of these "dimesons" gains a suppression factor $\sim 1/\sqrt{N}$ 
from each of the two groups, $N_r \sim N_s \sim N$ \cite{BD}.}
whereas the interactions between the baryons and mesons are not suppressed.
However, if there is a point $\xi_0$ in the space of parameters
$\{g_i\}$ where some state becomes massless then the usual large-$N$ 
picture can break down. 
These states are baryonic (or composites
of these) with a mass that generically goes
like $N$, but near the critical point $\xi_0$ one finds that  
$M\sim N|\xi-\xi_0|^\eta$ for some exponent $\eta$. 
In the double-scaling limit in which $N\to\infty$ and $\xi\to\xi_0$ such
that $M$ is fixed, the theory decomposes into two 
sectors ${\cal H}\to{\cal H}_-\times{\cal H}_+$. For the sector ${\cal H}_-$,  
the usual rules of the large-$N$ expansion are modified by diagrams 
involving the state of mass $M$ running around in loops. As described
in \cite{BD} there is now an effective string coupling as in \eqref{effg}.
The sector ${\cal H}_+$, on the other hand, decouples and is bound by
conventional large-$N$ reasoning (and hence is not interesting). 
What is remarkable is that there is
a natural candidate for the string dual to the sector ${\cal H}_-$, 
namely a non-critical string background
of the type introduced in \cite{GK}. These non-critical backgrounds
were originally studied as holographic duals to certain Little
String Theories (LST) and double-scaled LST
\cite{LST,holog,GKP,LSZLST}. 
In the low-energy limit these LST 
reduce to 4d ${\cal N}=2$ theories in the proximity of Argyres-Douglas
singularities which are non-trivial superconformal field theories 
\cite{AD,ARSW,EHIY}.
The fact that these backgrounds preserve ${\cal N}=2$ supersymmetry
seems to contradict the proposed duality.  Remarkably, the $F$-terms
of the theory are consistent with this supersymmetry enhancement 
in the ${\cal H}_-$ sector \cite{BD}.
It is interesting to try to extend this kind of duality and search for 
more general classes of double-scaling limits in supersymmetric gauge
theories. In this paper, we will achieve this aim and we
will show in many cases that these double-scaling limits exist and that
the Hilbert space of the theory 
still exhibits the above splitting into two decoupled sectors. As before,
the sector  ${\cal H}_-$ has non-trivial dynamics in the double-scaling limit
weighted by the effective string coupling \eqref{effg}. However, in the generic
situation we consider, this sector only has $\N=1$ supersymmetry
since the effects of the superpotential do not vanish in the limit. 
This is different from the case studied in \cite{BD},
where the superpotential vanished in the ${\cal H}_-$ sector, 
which is consistent with the
enhancement to $\N=2$ supersymmetry. 
Consequently, what is missing from our analysis is the precise identification of the 
dual string theory itself 
\footnote{A construction of four-dimensional ${\cal N}=1$
non-critical strings dual to ${\cal N}=1$ quiver gauge theories has recently 
appeared in \cite{Israel}}.
We will also find examples where the sector ${\cal H}_-$ does not contain
any $U(1)$ group but has a mass gap and light meson-like composite states.

So our focus is on large-$N$ double-scaling limits in supersymmetric gauge
theories. As we have argued, in order to have such a limit we need a
scenario in which there are states which we can tune to have a finite
mass as $N\to\infty$. The question 
is how can we verify that such double-scaling limits
actually exist because we do not generally have analytic control
over the masses of particles in an $\N=1$ theory? The answer is that,
under certain circumstances, we can see the effects of a nearly
massless particle in various $F$-terms and these can often be calculated exactly.
For instance, if the particle is charged under a $U(1)$ gauge
symmetry that survives in the IR, then we will see the effects of the
light particle in the associated renormalization of the $U(1)$
coupling. This
was one of the strategies employed in  \cite{BD} to explore these
double-scaling limits. In the present work, we will find situations
where the light states are not charged under any $U(1)$ in the IR. So
in order to ``see'' these kinds of states we will have to investigate 
other $F$-term couplings of the gauge theory. The idea is
to consider such couplings that arise when 
the gauge theory is coupled to supergravity. In that case
there is a whole set of gravitational $F$-terms \cite{AGNT,BCOV}. 
There are two series of terms that interest us 
\EQ{
\Gamma_1=\sum_{g=0}^\infty \int
d^4x\,d^2\theta\,(F_{\alpha\beta}F^{\alpha
\beta})^{g}\sum_{i=1}^{s}N_i\frac{\partial F_g(S)}{\partial S_i}
\label{fgft1}
}
and
\EQ{
\Gamma_2=\sum_{g=1}^\infty \int d^4x\,d^2\theta\,{\cal W}_{\alpha\beta
\gamma}{\cal
  W}^{\alpha\beta\gamma}(F_{\delta\epsilon}F^{\delta\epsilon})^{g-1}
F_g(S)\ .
\label{fgft2}
}
Here, ${\cal W}_{\alpha\beta\gamma}$ and $F_{\alpha\beta}$ are the
${\cal N}=2$ gravitino and gravi-photon superfields of the gravity
sector, respectively. The $F_g(S)$ are the gauge theory objects:
functions of the $s$ glueball superfields $\{S_i\}$. Note that the $g=0$
term of $\Gamma_1$ is the ordinary glueball superpotential:
\EQ{
W_\text{gb}=\sum_{i=1}^{s}N_i\frac{\partial
  F_0(S)}{\partial S_i}\ ,
\label{gbsp}
}
which determines the vacua of the theory.\footnote{We are assuming
  that the VEVs of the supergravity fields vanish otherwise we have to
  keep all the terms in $\Gamma_1$ in the superpotential as in
  \cite{Ooguri:2003qp}. }

In the usual large-$N$ limit, the contribution from a genus $g$ graph
is suppressed by 
$N^{2-2g}$. However, if there is a scenario as described above where a
state is becoming massless then this suppression can be 
subverted. The effect of such a state on $F_g$ can be crudely
estimated by considering the light state propagating around a 1-loop
graph with $g$ insertions \cite{BD,Vafaconifold}. 
This suggests $F_g\sim M^{2-2g}$, which together with the identification
$M\sim\Delta$ means that the gravitational $F$-term $F_g$ behaves just
like the genus $g$ free energy of a matrix model as in \eqref{fdsl}. 
The conclusion is that the quantities $F_g$ appearing in the 
gravitational $F$-terms provide another signal of a double-scaling limit.
Of course it is no accident
that these couplings behave like the free energy of a matrix model
since they can be calculated by just such a matrix model as described
by Dijkgraaf and Vafa \cite{Dijkgraaf:2002fc}. This 
matrix model is defined in terms of an $\hat N\times\hat N$ matrix
$\hat\Phi$ ($\hat N$ here is not to be confused with $N$ of the field
theory) with the bare superpotential in \eqref{bsup}:
\EQ{
\exp\sum_{g=0}^\infty F_g g_s^{2g-2}=\int
d\hat\Phi\,\exp\Big(-g_s^{-1}\text{Tr}\,W(\hat\Phi)\Big)\ .
\label{L_DV}
}
The integral is to be understood as a saddle-point expansion around a
critical point where $\hat N_i$ of the eigenvalues sit in the critical
point $a_i$ (but only if the associated $N_i>0$) and by definition the
glueball superfields are identified with the quantities 
\EQ{
S_i=g_s\hat N_i\ ,\qquad S=\sum_{i=1}^{s}S_i=g_s\hat N
\label{defs}
}
in the matrix model. It is worth remarking that in the matrix model we
work at large $\hat N$ and that this procedure would yield exact results
for any value of $N$. But, of course, 
we also want to take $N$ to be large in the double-scaling limit.

It should now be apparent that the double-scaling limit of the
gravitational $F$-terms \eqref{fgft1}\eqref{fgft2} in the gauge theory corresponds
precisely to the notion of a double-scaling limit in the ``old''
matrix model and in certain cases we 
can use standard matrix model technology to investigate the phenomenon. 

The results presented in this paper have some overlap with earlier
work. The idea that the standard large-$N$ expansion breaks down near the
critical points in the moduli/parameter space of four-dimensional
gauge theories has been emphasized in a series of papers by Ferrari 
\cite{ferr1,ferr2,ferr3}, 
which also contain proposals for double-scaling limits. 
The four-dimensional interpretation of
double-scaling in the Dijkgraaf-Vafa matrix model has also been
discussed before in \cite{Dijkgraaf:2002fc,ferr1}. 

The plan of the paper is as follows. 
In section \ref{Engineer}, we will give a general description
of double-scaling limits and explain how they can be realized on-shell. 
In section \ref{DoubleFterms}, we will study the exact $F$-term couplings 
of the various models and deduce that in the double-scaling limit 
the Hilbert space of theory decomposes into two decoupled sectors
${\cal H}_+$ and ${\cal H}_-$, where ${\cal H}_-$ keeps finite
interactions in the limit.   
In section \ref{DoubleFg}, we will analyze the behaviour of the higher
genus terms of the matrix model free energy $F_g$ at the critical point
and show that $F_g \sim \Delta^{2-2g}$.
In section \ref{PhysicsDoublePoints}, we will discuss in detail 
the case where the near-critical curve has double points.    
We will argue that in the critical limit where a double point collides
with a branch point a meson-like state, which can be thought of as
the bound state of two dibaryons, becomes massless.

\section{Engineering Double-Scaling Limits in the Matrix Model}
\label{Engineer}

The central object in matrix model theory is the resolvent 
\EQ{
\omega(x)=\frac1{\hat N}\text{Tr}\,\frac1{x-\hat\Phi}\ .
}
At leading order in the $1/\hat N$ expansion, $\omega(x)$ is valued on
the spectral curve $\Sigma$, a hyper-elliptic Riemann surface
\EQ{
y^2=\frac1{(N\varepsilon)^2}\big(W'(x)^2+f_{\ell-1}(x)\big)\ .
\label{mmc}
}
The numerical prefactor is chosen for convenience. In terms of this curve
\EQ{
\omega(x)=\frac1{2S}\big(W'(x)-N\varepsilon y(x)\big)\ .
}
In \eqref{mmc}, $f_{\ell-1}(x)$ is a polynomial of order $\ell-1$
whose $\ell$ coefficients are moduli that are determined by the
$S_i$. In general, the spectral curve can be viewed as a double-cover
of the complex plane connected by $\ell$ cuts. For the saddle-point of
interest only $s$ of the cuts are opened and so only $s$ of
the moduli $f_{\ell-1}(x)$ can vary. Consequently $y(x)$ has $2s$
branch points and $\ell-s$ zeros:\footnote{Occasionally, for clarity, 
we indicate the order of a polynomial by a subscript.}
\EQ{
\Sigma:\qquad y^2=Z_m(x)^2\sigma_{2s}(x)
\label{mc}
}
where $\ell=m+s$ and 
\EQ{
Z_m(x)=\prod_{j=1}^m(x-z_j)\ ,\qquad
\sigma_{2s}(x)=\prod_{j=1}^{2s}(x-\sigma_j)\ .
}
The remaining moduli are related to the
$s$ parameters $\{S_i\}$ by \eqref{defs}
\EQ{
S_i=g_s\hat N_i=N\varepsilon\oint_{A_i}y\,dx\ ,
}
where the cycle $A_i$ encircles the cut which opens out around the
critical point $a_i$ of $W(x)$. 

Experience with the old matrix model teaches us that double-scaling
limits can exist when the parameters in the potential are varied in
such a way that combinations of branch and double points come
together. In the neighbourhood of such a critical point,\footnote{We
  have chosen for convenience to take all the double zeros $\{z_j\}$ 
into the critical region.} 
\EQ{
y^2\longrightarrow C Z_m(x)^2B_n(x)\ ,\qquad
B_n(x)=\prod_{i=1}^n(x-b_i)\ ,
\label{redc}
}
where $z_j,b_i\to x_0$, which we can take, without loss of generality, 
to be $x_0=0$. The double-scaling limit involves first taking $a\to0$
\EQ{ 
x=a\tilde x\ ,\qquad z_i=a\tilde z_i\ ,\qquad b_j=a\tilde b_j
\label{dsl}
}
while keeping tilded quantities fixed.
In the limit, we can define the near-critical curve
$\Sigma_-$:\footnote{For polynomials, we use the notation
  $\tilde f(\tilde x)=\prod_i(\tilde
  x-\tilde f_i)$, where $f(x)=\prod_i(x-f_i)$, 
$x=a\tilde x$ and $f_i=a\tilde f_i$.}
\EQ{
\Sigma_-:\qquad y_-^2=\tilde Z_m(\tilde x)^2\tilde B_n(\tilde
x)\ .
\label{ncc}
}
We will argue, generalizing \cite{BD}, that in the limit $a\to0$, in
its sense as a complex manifold, the curve $\Sigma$ factorizes as
$\Sigma_-\cup\Sigma_+$. The complement to the near-critical
curve is of the form
\EQ{
\Sigma_+:\qquad y_+^2=x^{2m+n}C_{2s-n}(x)\ .
}
where $C_{2s-n}(x)$ is regular at $a=0$.

In the $a\to0$ limit, we will show in Section \ref{DoubleFg}  that the
genus $g$ free energy gets a dominant contribution from $\Sigma_-$
of the form
\EQ{
F_g\thicksim \big(Na^{(m+n/2+1)}\big)^{2-2g}\ .
\label{dsp}
}
Note that in this equation $N$ is the one from the field theory and
not the matrix model $\hat N$. This motivates us to define the 
double-scaling limit
\begin{equation}
a \rightarrow 0 \ , \qquad N \rightarrow \infty
\ , \qquad \Delta \equiv N a^{m+n/2+1} = \text{const} \ .
\label{limitI1}\end{equation}
Moreover, the most singular terms in $a$ in \eqref{dsp} depend only on the
near-critical curve \eqref{ncc} in a universal way.

\subsection{Engineering the double-scaling limit on-shell}

However, there is still an important issue to address. In the context
of supersymmetric gauge theories, the moduli $\{S_i\}$ are fixed by
extremizing the glueball superpotential \eqref{gbsp}. It is not, {\it
  a priori\/}, clear whether a double-scaling limit can be reached
whilst simultaneously being on-shell with respect to the glueball
superpotential. We now address this issue and show that
suitable choices of the coupling constants $\{g_i\}$ do indeed allow
for a double-scaling limit on-shell with respect to the glueball
superpotential. In general, the
potentials required are non-minimal. However, this is irrelevant for
extracting the universal behaviour that only depends on the
near-critical curve \eqref{redc}.

So the problem before us is to show that the critical point can be
reached simultaneously with being at a critical point of the
glueball superpotential. It is rather difficult to find
the critical points of the latter directly. 
Fortunately another more tractable method consists of comparing
the matrix model spectral curve \eqref{mmc}, the ``$\N=1$ curve'', with
the Seiberg-Witten curve of the underlying $\N=2$ theory that results when the
potential vanishes. The latter has the form
\EQ{
y_\text{SW}^2=P_N(x)^2-4\Lambda^{2N}\ ,
}
where $P_N(x)=\prod_{i=1}^N(x-\phi_i)$. Here, $\{\phi_i\}$ are a set of
coordinates on the Coulomb branch of the $\N=2$ theory 
and $\Lambda$ is the usual scale of strong-coupling effects in the
$\N=2$ theory.

When the $\N=2$ theory is deformed by addition of the superpotential
\eqref{bsup}, it can be shown that a vacuum exists when the Seiberg-Witten
curve and the $\N=1$ curve represent the same underlying 
Riemann surface \cite{CSW,deBoer:1997ap,CIV}. 
In concrete terms this means that, on-shell,
\SP{
&y_\text{SW}^2=P_N(x)^2-4\Lambda^{2N}=H_{N-s}(x)^2\sigma_{2s}(x)\\
&y^2=\frac1{(N\varepsilon)^2}\big(W^{\prime}(x)^2
+f_{s+m-1}(x)\big)=Z_m(x)^2\sigma_{2s}(x)\ ,
\label{con}
}
In these equations,  $H_{N-s}(x)$, $\sigma_{2s}(x)$, $Z_{m}(x)$
are polynomials of the indicated order, and we choose
(in order to remove some redundancies)
\EQ{
H_{N-s}(x)=x^{N-s}+\cdots\ ,\qquad
\sigma_{2s}(x)=x^{2s}+\cdots,\qquad
Z_m(x)=x^m+\cdots\ .
}
Both curves describe the same underlying Riemann surface,
namely the reduced curve of genus $s-1$ which is a hyper-elliptic
double-cover of the complex plane with $s$ cuts.
All-in-all there are $2(N+l)$ equations for the same number of unknowns in
$\{P,H,\sigma,Z,f\}$. There are many solutions to these equations
and we can make contact with the description of the vacua in Section 1
by taking the classical limit $\Lambda\to0$; whence
\EQ{
P_N(x)\to\prod_{i=1}^{\ell}(x-a_i)^{N_i}\ ,\qquad\sum_{i=1}^{\ell}N_i=N\ ,
}
so $N_i$ of the eigenvalues of the Higgs field classically lie at the
critical point $a_i$ of $W(x)$. Quantum effects then have the effect
of opening the points $a_i$ into cuts (if $N_i>0$). The number of
$N_i>0$, {\it i.e\/}~the number of cuts, is equal to $s=\ell-m$.

We now to turn to explicit solutions of \eqref{con}. The method we
shall adopt is to first find solutions for a $U(p)$ gauge theory and
then apply the ``multiplication by $N/p$ map'' \cite{CIV}, with $N/p$ integer. 
This will yield a solution for a $U(N)$ gauge group and will allow to take a large $N$
limit with $p$ fixed.

\subsection{No double points}

We now describe how to engineer 
the case where the near critical curve \eqref{redc} has 
no double points, so $m=0$. This is the situation considered in 
\cite{BD,Eguchi:2003wv,bert}. 
In this case, we first consider the consistency conditions
\eqref{con} for a $U(p=n)$ gauge theory with $W(x)$ of order
$\ell=n+1$. In this case, \eqref{con} are trivially satisfied with
\EQ{
W'(x)=N\varepsilon P_n(x)\ ,\qquad f_{n-1}(x)=-4N^2\varepsilon^2\Lambda^{2n}\ .
}
Notice that with our minimal choice of potential, the on-shell curve
actually implies that $S=0$ since the coefficient of $x^{n-1}$ in
$f_{n-1}(x)$ vanishes and so the resolvent falls faster than $1/x$ at
infinity. This, of course, is pathological from the
point-of-view of the old matrix model and may be remedied by using a
non-minimal potential with extra branch points or double points
outside the critical region. However, in the holomorphic context in
which we are working, having $S=0$ is perfectly acceptable and we
stick with it.  
The on-shell curve consists of an $n$-cut hyperelliptic curve and one
can verify, by taking the classical limit, that $N_i=1$, $i=1,\ldots,n$. The
double-scaling limit involves a situation where $n$ branch points, one
from each of the cuts, come together. This can be arranged by having
\EQ{
W'(x)=N\varepsilon\big( B_n(x)
+2\Lambda^n\big)\ ,\qquad B_n(x)=\prod_{j=1}^n(x-b_j)
\label{cioppino}}
and then taking the limit \eqref{dsl}. In this case, the near-critical
curve $\Sigma_-$ \eqref{ncc} is of the form
\EQ{
y_-^2=\tilde B_n(\tilde x)\ .
}
The important point is that we can tune to the critical region whilst
keeping the theory on-shell with respect to the glueball
superpotential by simply changing the parameters $\{b_j\}$ which
appear in the potential.

Now that we have found a suitable vacuum of a $U(n)$ theory, we now
lift this to a $U(N)$ theory with the multiplication by $N/n$ map
\cite{CIV}. 
Under this map, the $\N=1$ curve remains intact, including the 
potential $W(x)$ whilst the Seiberg-Witten curve of the $U(N)$ theory is 
\EQ{
y_{SW}^2=P_N(x)^2-4\Lambda^{2N}=\Lambda^{2(N-n)}{\cal
  U}_{\tfrac Nn-1}\Big(\frac{P_n(x)}{2\Lambda^n}\Big)^2\big(P_n(x)^2-
4\Lambda^{2n}\big)\ ,
\label{mnm}
}
where ${\cal U}_{\tfrac Nn-1}(x)$ is a Chebishev polynomial of the second
kind. The vacuum of the $U(N)$ theory has $N_i=N/n$, $i=1,\ldots,n$. 

Notice that in the near critical region the Seiberg-Witten curve 
is identical to $\Sigma_-$, up to a rescaling:
\EQ{
y_\text{SW}^2\longrightarrow\Big(\frac{2N}{n}\Big)^2
\Lambda^{2N-n}B_n(x)\ .
\label{nhy}
}
This is
simply a reflection of the observation of \cite{BD} that the decoupled
sector has enhanced $\N=2$ supersymmetry. Moreover, if $C$ is a cycle
which is vanishing as $a\to0$ then the integral of the Seiberg-Witten
differential around $C$, which gives the mass of a BPS state carrying
electric and magnetic charges in the theory, becomes
\EQ{
\oint_C\frac{xP'_N(x)\,dx}{y_\text{SW}}
\longrightarrow - \Lambda^{-n/2} Na^{n/2+1}\oint_C y_-\,d\tilde x\ .
}
Notice that in the double-scaling limit \eqref{limitI1} (with $m=0$)
the mass of the state is fixed. This state is a dibaryon that carries electric 
and magnetic charges of the IR gauge group. In the
double-scaling limit, therefore, a set of mutually non-local dibaryons
become very light.\footnote{For $n=2$ there is only a single light dibaryon.}
In fact, the Seiberg-Witten curve at the critical point, $a=0$, has the form
\EQ{
y_\text{SW}^2=4\Big(\frac{N}{n}\Big)^2
\Lambda^{2N-n}x^{n}\ ,
}
which describes a ${\bf Z}_n$ or $A_{n-1}$ Argyres-Douglas singularity 
\cite{AD,ARSW,EHIY}.

\subsection{With double points}

For the case with double points, we cannot simply take two of the branch points
$\{b_j\}$ in \eqref{cioppino} above to be the same. If we simply did
that then the zero of the Seiberg-Witten curve, by which we mean a 
zero of the polynomial $H_{N-s}$ in \eqref{con}, would also be a zero 
of the $\N=1$ curve as well. By the analysis of \cite{deBoer:1997ap},
this would imply that the condensate of the associated massless dibaryon
would be vanishing. 
On the contrary, if the zero of the Seiberg-Witten curve were not a zero of the 
$\N=1$ curve, the putative massless dibaryon would be condensed
and the dual $U(1)$ would be confined.  
We need to arrange the situation so that any zero of the Seiberg-Witten
curve is not simultaneously a zero of the $\N=1$ curve. 

A suitable $\N=1$ curve which reduces to \eqref{redc} in the near-critical
region is
\EQ{
y^2=Z_m(x)^2B_n(x)\big(B_n(x)H_{r}(x)^2+4\Lambda^{2r+n}\big)\ .
\label{noc}
}
In this case, we have $\ell=m+n+r$, $s=n+r$ and
\EQ{
W'(x)=N\varepsilon Z_m(x)B_n(x)H_{r}(x)\ ,\qquad
f_{\ell-1}(x)=4N^2\varepsilon^2\Lambda^{2r+n}Z_m(x)^2B_n(x)\ .
}
Notice that in order that $f_{\ell-1}(x)$ has order less than $\ell$ we
require $r>m$. The curve \eqref{noc} is actually on-shell with respect
to the Seiberg-Witten curve of a $U(2r+n)$ theory with
\EQ{
P_{2r+n}(x)=H_{r}(x)^2B_n(x)+2\Lambda^{2r+n}\ .
}
In the classical limit, we have two eigenvalues at each of the
zeros of $H_r(x)$ and one in each of the zeros of $B_n(x)$.
Once again we can employ the multiplication map \eqref{mnm} (with $n$
replaced by $2r+n$) to find the vacuum of the $U(N)$ theory we are after. 

Notice that the double points of the Seiberg-Witten curve $\{h_i\}$
are not generally zeros of the curve \eqref{noc}, which means that the
associated dyons are condensed.
The near-critical curve $\Sigma_-$ in this case is
\EQ{
y_-^2=\tilde Z_m(\tilde x)^2\tilde B_n(\tilde x) \ ,
}
while in the near-critical region the Seiberg-Witten curve becomes
\EQ{
y_\text{SW}^2\longrightarrow4\Big(\frac{N}{n+2r}\Big)^2
\Lambda^{2N-2r-n}H_r(0)^2B_n(x)\ .
\label{swnc}
}
where we assumed that the zeros of $H_r(x)$ lie outside the critical
region. In this case, the integral of the Seiberg-Witten differential
around a vanishing cycle diverges in the double-scaling limit:
\EQ{
\oint_C\frac{xP'_N(x)\,dx}{y_\text{SW}}\thicksim Na^{n/2+1}=\Delta
a^{-m}\to\infty\ .
}
So in contrast to the case with no double points, 
the dibaryon states are very heavy. In addition, the
dyon condensate associated to the zero $h_i$ of $H_r(x)$ is given by an 
exact formula \cite{deBoer:1997ap} 
\EQ{
\langle m_i\tilde m_i\rangle=N\varepsilon y(h_i)\thicksim N
\to\infty\ ,
}
where we have assumed that $h_i$ stays fixed as $a\to0$.
So in the double-scaling limit the value of the condensate and hence
the confinement scale in the dual $U(1)$, or string tension, 
occurs at a very high mass scale. 

We now have a puzzle. How can there be an interesting double-scaling
limit in the gauge theory in this case 
if there are no light dibaryon as in the previous
example? The answer is that there are other light mesonic states in the theory
with a mass $\sim\Delta$ that we will identify in Section \ref{PhysicsDoublePoints}.

Notice that contrary to our choice above, if we scale $h_i\to0$ as $a\to0$ then
the tensions of the confining strings vanish and the 
theory is at an $\N=1$ superconformal fixed point in the infra-red
corresponding to one of the $\N=1$ Argyres-Douglas-type singularities
described in \cite{Eguchi:2003wv}. As the double points of 
the Seiberg-Witten curve $h_i$
move away from $0$ the associated dyons condense and the superconformal
invariance is spontaneously broken. The resulting Goldstone modes will
play an important r\^ole in the ensuing story and resolve the puzzle
alluded to above.

\section{The Double-Scaling Limit of $F$-Terms}\label{DoubleFterms}

Before we consider the $a\to0$ limit of the free energy, it is
useful to the consider this limit for other $F$-terms in the
low-energy effective action. 
The effective action is written in terms of chiral superfields $S_{l}$
and $w_{\alpha l}$ which are defined as gauge-invariant single-trace  
operators \cite{CDSW}
\SP{
S_{l} & = - \frac{1}{2\pi i}\oint _{A_{l}} \, dx\, \frac{1}{32\pi^{2}} 
{\rm Tr}_{N}\left[ \frac{W_{\alpha}W^{\alpha}}{x- \Phi}\right]\ ,\\ 
w_{\alpha l} & = \frac{1}{2\pi i}\oint _{A_{l}} \, dx 
\, \frac{1}{4\pi} {\rm Tr}_{N}\left[ 
\frac{W_{\alpha}}{x- \Phi}\right] \ . 
\label{gidef}
}
It will also be convenient to define component fields for each of
these superfields, 
\EQ{
S_{l}= s_{l}+\theta_{\alpha}\chi^{\alpha}_{l}+ \cdots \ ,\qquad
w_{\alpha l}=\lambda_{\alpha l} +\theta_{\beta}f^{\beta}_{\alpha
  l}+ \cdots \ .
\label{comp}
}
The component fields,  $s_{l}$ and $f_{l}$ 
are bosonic single trace operators whilst 
$\chi_{l}$ and $\lambda_{l}$ are fermionic single trace operators. 
In the large-$N$ limit, these operators should 
create bosonic and fermionic
colour-singlet single particle states respectively. 
It is instructive to consider the interaction vertices for these
fields contained in the $F$-term effective action whose general form is
given by  \cite{Dijkgraaf:2002fc,DV3,DVPW}
\begin{equation} 
{\cal L}_{F}=
{\rm Im}\left[\int d^{2}\theta \,\big( W_\text{gb} +
  W_\text{eff}^{(2)}\big)\right]
\label{fterm0}
\end{equation}
where
\EQ{
W_\text{eff}^{(2)}  = 
\frac{1}{2}\sum_{k,l} 
\frac{\partial^{2}{F_0}}{ \partial S_{k} \partial S_{l}}
\,w_{\alpha k}w^{\alpha}_{l} \ .
\label{fterm1}
}
Expanding (\ref{fterm0}) in components on-shell, we find terms like  
\begin{equation} 
\int\, d^{2}\theta\, W_\text{eff}^{(2)} \supset 
V^{(2)}_{ij}f^{i}_{\alpha\beta}f^{\alpha\beta j}  + 
V^{(3)}_{ijk} \chi^{i}_{\alpha}f^{\alpha\beta j}\lambda^{k}_{\beta}
+V^{(4)}_{ijkl}\chi^{i}_{\alpha}\chi^{\alpha j}
\lambda^{k}_{\beta}\lambda^{\beta l}\ ,
\label{vijk}
\end{equation}
where
\begin{equation}
V^{(L)}_{i_{1}i_{2}\ldots i_{L}} =
\frac{\partial^{L} {F_0}}{\partial S_{i_{1}} \partial S_{i_{2}}
\ldots\partial S_{i_{L}}}
\label{vp}
\end{equation}
for $L=2,3,4$. 
In the large-$N$ limit, $V^{(L)}$ scales like $N^{2-L}$.
We will also consider the $2$-point vertex coming from the
glueball superpotential
\begin{equation}
\int\, d^{2}\theta\, W_\text{gb} \supset 
H^{(2)}_{ij}\chi_{\alpha}^{i}\chi^{\alpha j}\ ,
\end{equation}
where 
\begin{equation}
H^{(2)}_{ij}= 
\frac{\partial^{2} W_\text{gb} }{\partial S_{i} \partial S_{j}}\ . 
\end{equation}
The matrix $H^{(2)}_{ij}$ therefore 
effectively determines the masses of the chiral multiplets $S_{l}$.   
Note that, in the large-$N$ limit, $H^{(2)}$ scales like $N^0$.

We begin by considering the couplings $V^{(2)}_{ij}$ of the low-energy $U(1)^s$ gauge
group. Each of the $U(1)$'s is associated to one of the glueball fields
$S_i$, or equivalently the set of 1-cycles $\{A_i\}$ on $\Sigma$. 
If we ignore the $U(1)$ associated to the overall 't Hooft coupling
$S$, or the cycle $A_\infty=\sum_{i=1}^sA_i$ which can be pulled off to
infinity, the couplings of the remaining ones are simply the elements of the
period matrix of $\Sigma$. In order to take the $a\to0$ limit, it is
useful to choose a new basis of 1-cycles  $\{\tilde A_i,\tilde B_i\}$,
$i=1,\ldots,s-1$, which is specifically adapted to the
factorization $\Sigma\to\Sigma_-\cup\Sigma_+$. 
The subset of cycles with $i=1,\ldots,[n/2]$  
vanish at the critical point while the cycles $i=[n/2]+1,\ldots,s-1$ 
are the remaining cycles which have zero intersection with
all the vanishing cycles.  

If we define the periods on $\Sigma$
\EQ{
M_{ij} = \oint_{\tilde B_j} 
\frac{x^{i-1}}{\sqrt{\sigma(x)}}\, dx 
\ ,
\qquad
N_{ij} =
\oint_{\tilde A_j} \frac{x^{i-1}}{\sqrt{\sigma(x)}}\, dx \ ,
\label{MijNij}
}
then the period matrix, in this basis, is simply
\EQ{
\Pi=N^{-1}M\ .
}
In the Appendix, we calculate the $a\to0$ limit of these matrices.
The results are summarized in \eqref{lim} and \eqref{NinvLimit}. 
Using these results, we have
\begin{equation}
\Pi\longrightarrow
\left(
\begin{array}{cc}
N^{-1}_{--} M_{--} & \quad N_{--}^{-1} M_{-+}^{(0)}
+ {\cal N} M_{++}^{(0)}
\\
0 & \quad \big( N^{(0)}_{++} \big)^{-1} M_{++}^{(0)} \\
\end{array} 
\right) \ .
\label{Pi1}\end{equation}
Let us look more closely at the structure of each
block in the above matrix. First of all, by (\ref{N--}) 
\begin{equation}
\big( N_{--}\big)_{ij} 
\thicksim 
a^{ n/2 - j }
\, f^{(N)}_{ij}( \tilde b_l) \ ,\qquad \big(M_{--}\big)_{ij} 
\thicksim 
a^{ n/2 - j }
\, f_{ij}^{(M)}( \tilde b_l) \ ,
\label{Ninv--}\end{equation}
which implies  
\begin{equation}
\big( N_{--} \big)^{-1}_{ij} \, 
\big( M_{--} \big)_{jk}
= 
f^{(N)-1}_{ij} ( \tilde b_l ) 
f^{(M)}_{jk}( \tilde b_l) 
= \Pi^{-}_{ik} ( \tilde b_l)
\ .
\label{Piminus}\end{equation}
Furthermore, since $N_{--}^{-1}$ vanishes 
in the limit $a \to 0$, we find that
\begin{equation}
N_{--}^{-1} M_{-+}^{(0)}
+ {\cal N} M_{++}^{(0)} =
N_{--}^{-1} \,M_{-+}^{(0)}
- N_{--}^{-1} \,N_{-+}^{(0)} 
\big( N_{++}^{(0)} \big)^{-1} 
M_{++}^{(0)} \rightarrow 0 \ .
\end{equation}
Therefore, the period matrix has the following block-diagonal form
in the double-scaling limit
\begin{equation}
\Pi \longrightarrow 
\left(
\begin{array}{cc}
\Pi^- & 0
\\
0 & \Pi^+ \\
\end{array} 
\right) \ .
\label{Pifinal}\end{equation} 
The upper block $\Pi^-$ is actually
the period matrix of the near-critical spectral curve $\Sigma_-$ 
\eqref{ncc} since the cycles 
$\{\tilde A_i,\tilde B_i\}$, for $i\leq[n/2]$ 
form a standard homology basis for $\Sigma_-$. Similarly, the
lower block $\Pi^+$ is the period matrix of $\Sigma_+$. So in
the limit $a\to0$ the curve $\Sigma$ factorizes as
$\Sigma_-\cup\Sigma_+$. The fact that the
period matrix factorizes is evidence of the more stringent claim that
the whole theory consists of two decoupled sectors ${\cal
  H}_-$ and ${\cal H}_+$ in the double-scaling limit.
Note that although we did not consider it, 
the $U(1)$ associated to $S$ only couples to the ${\cal H}_+$ sector.

We can extend this discussion to include other $F$-terms that are
derived from the glueball superpotential. 
For example,  consider the 3-point vertex
\EQ{
V^{(3)}_{ijk} = 
\frac{\partial^3 F_0}{\partial\tilde S_i \partial\tilde S_j
  \partial\tilde S_k}\ .
}
Here, the $\tilde S_i$ as defined as in
\eqref{defs} but with respect to the cycles $\tilde A_i$. They are
related to the $S_i$ by an electro-magnetic duality transformation.
There is a closed expression for these couplings 
of the form \cite{CMMV,Krichever,DVWDVV}
\begin{equation}
V^{(3)}_{ijk} 
= \frac1{N\varepsilon}\sum_{l=1}^{2s}\text{Res}_{b_l} 
\ \frac{\omega_i \,\omega_j \,\omega_k}{dx dy} \  ,
\label{d3FdS3}\end{equation}
where $\{\omega_j\}$ are the holomorphic 1-forms normalized with
respect to the basis $\{\tilde A_i,\tilde B_i\}$.
So we can deduce the behaviour of the couplings
from our knowledge of the scaling of $\omega_j$. This is derived in
the Appendix. We find that the couplings are regular as $a\to0$,
except if $i,j,k\leq[n/2]$ in which case,
\EQ{
V^{(3)}_{ijk} \longrightarrow
\big(N \varepsilon a^{m+n/2+1} \big)^{-1} 
\sum_{l=1}^{n} \text{Res}_{\tilde{\sigma}_l} 
\ \frac{\tilde{\omega}_i \,\tilde{\omega}_j 
\,\tilde{\omega}_k}{d\tilde{x} dy_-} \ ,
\label{ints}
}
where the $\{\tilde{\omega}_i\}$ are the one-forms on $\Sigma_-$.
Therefore, in the double-scaling limit proposed in \eqref{limitI1}, we
find that these interactions remain finite $\sim\Delta^{-1}$, while the
other 3-point vertices $\to0$. This is yet further  
evidence of the decoupling of the Hilbert space into
two decoupled sectors where the
interactions in the ${\cal H}_-$ sector remain finite in the
double-scaling limit while those in ${\cal H}_+$ go to zero.
Notice, also that these interactions of the ${\cal H}_-$ sector 
depend universally on $\Sigma_-$.

The final $F$-term quantity that we consider is the 
Hessian matrix for the glueball superfields 
\EQ{
H^{(2)}_{jk} = 
\frac{\partial^2 W_\text{gb}}{\partial\tilde S_j 
\partial\tilde S_k} \ .
}
Using (\ref{d3FdS3}) we find 
\EQ{
H^{(2)}_{jk} = \sum_{i=1}^s N_i 
\frac{ \partial^3 F_0 }{\partial \tilde S_i\partial\tilde S_j 
\partial\tilde S_k}  
=\frac1{N\varepsilon} \sum_{l=1}^{2s} \text{Res}_{b_l} \ 
\frac{T\,\omega_j \,\omega_k}{dx dy} \ ,
\label{hess}
}
where we have defined the 1-form $T$ 
\EQ{
T=N\varepsilon\sum_{i=1}^s\frac{\partial y\,dx}{\partial S_i}\ .
}
It is known that $T$ can be can be written simply in terms of the
on-shell Seiberg-Witten curve \cite{CSWII}:
\EQ{
T= d \log(P_N+y_\text{SW})\ .
\label{Tdxexplicit}
}
In the limit $a\to0$, we can take the near-critical expressions for 
$y_\text{SW}$ in \eqref{swnc} and for $P_N(x)=2\Lambda^N$ to get the
behaviour 
\EQ{
T\longrightarrow\Lambda^{-r-n/2}H_r(0)\frac
N{n+2r}a^{n/2}d\sqrt{\tilde B(\tilde x)}\thicksim Na^{n/2}\ .
}
We also need
\EQ{
dy\longrightarrow a^{m+n/2}d\big(\tilde 
Z_m(x)\sqrt{\tilde B(\tilde x)}\big)\thicksim a^{m+n/2}\ .
} 
The scaling of the holomorphic differentials is determined in the
Appendix.

Counting the powers of $N$ and $a$, we find that for any $j$ and
$k$, $H^{(2)}_{jk}$ goes like an inverse power of $a$ and hence
diverges in the double-scaling limit (the powers of $N$ cancel). This,
however, presents us with a puzzle. In the case without double points
described in \cite{BD}, the Hessian was shown to vanish for the
${\cal H}_-$ sector, {\it i.e.\/}~$j,k\leq[n/2]$. Let us see how this is
compatible with the scaling we have just seen.
In the case, $j,k \leq [n/2]$,
\EQ{
H^{(2)}_{jk} \thicksim a^{-(m+1)} 
\sum_{l=1}^n \text{Res}_{\tilde{b}_l} \ \Big[
\frac{ d\sqrt{\tilde{B}_{n}(\tilde{x})}
\ \tilde{\omega}_j \,\tilde{\omega}_k}
{d\tilde{x} \ d \Big( \tilde{Z}_m(\tilde{x}) \,
    \sqrt{\tilde{B}_n(\tilde{x})} \Big)} 
\Big]\  ,
\label{H2--}
}
where 
\EQ{
 \tilde{\omega}_j = \frac{ \tilde{L}_j(\tilde{x})}{ \sqrt{ \tilde{B}_n(\tilde{x}) }}
 d \tilde{x} .
} 
and $\tilde{L}_j(\tilde{x})$ is a  polynomial of degree $[n/2]-1$.
Note that the differential $\tilde{\omega}_j \,\tilde{\omega}_k/d\tilde{x}$
has simple poles at $\tilde{x}=\tilde{b}_l$ on the curve $\Sigma_-$:
\EQ{
\frac{\tilde{\omega}_j \,\tilde{\omega}_k}{d\tilde{x}} = 
\frac{\tilde{L}_j(\tilde{x})
  \,\tilde{L}_k(\tilde{x})}{\tilde{B}_n(\tilde{x})}d\tilde{x} \ , 
}
but has no pole at $\tilde{x}=\infty$. For example for $n$ odd, we find  
\EQ{
\frac{\tilde{\omega}_j \,\tilde{\omega}_k}{d\tilde{x}} 
\longrightarrow \frac{d \tilde{x}}{\tilde{x}^3} \ .
}
This means that in the case with no double points, $m=0$, the 
Hessian matrix elements (\ref{H2--}) vanish identically:
\EQ{
H^{(2)}_{jk} \thicksim a^{-1} 
\sum_{l=1}^n \text{Res}_{\tilde{b}_l} \ \Big[
\frac{ \tilde{\omega}_j \,\tilde{\omega}_k}
{d\tilde{x}} \Big]
= 0 \ ,
}
because the sum of all residues of a meromorphic differential
on the compact near-critical curve $\Sigma_-$ is identically zero. 
This is precisely the result found in \cite{BD}. 
On the other hand, if $m > 0$, the Hessian matrix element 
will not vanish in general, because the differential 
on the right-hand side of (\ref{H2--}) has extra simple poles 
at the roots of
\EQ{
2 \tilde{Z}'_m(\tilde{x}) \tilde{B}_n(\tilde{x}) 
+ \tilde{Z}_m(\tilde{x}) \tilde{B}'_n(\tilde{x}) = 0 \ .
}

This result is very significant because it highlights an important
difference between the case with and without double points.
Even though we do not have control over the kinetic terms of the
glueball states, we take this behaviour of the Hessian matrix
to signal that, with double points, 
the masses of the glueball fields 
become very large in the double-scaling limit.
This is to be contrasted with the case without double points 
studied in \cite{BD},
where the appearance of the $[n/2]$ massless glueballs was interpreted as
evidence that supersymmetry is enhanced to ${\cal N}=2$
in the double-scaling limit. 

\section{The Double-Scaling Limit of the Free Energy}\label{DoubleFg}

In this section, we will consider the behaviour of the free energy in
the limit $a\to0$.
The most powerful methods for calculating the $F_g$ involves
using orthogonal polynomials which we review in Section 4.2; however,
as we shall see, these techniques appear only to be successful for the
cases with one or two branch points and any
additional number of double points.
The only known way
to calculate the $F_g$ in general involves analyzing the loop
equations and in particular using the 
algorithms recently developed in \cite{Eynard,ChekhovEynard}. We use
these techniques to prove our scaling ansatz \eqref{limitI1} in Section 4.3.
Before this, it is useful to consider a situation 
which is exactly solvable. This is
the case when there are two branch points and no double points: $n=2$
and $m=0$. 

\subsection{Simple example: two branch points}

In this section, we consider in some detail the case when the critical
point involves the collision of two branch points. Most of the
formulae that we use are taken from \cite{DGKV}. With two branch points
and no double points, we
can engineer as in Section \ref{Engineer} with a curve of the form
\EQ{
y^2=(x^2-a^2)(x^2-b^2)
}
in the limit $a\to0$. In order to be on-shell, we require
\EQ{
b^2=a^2+4\Lambda^2
}
because then 
\EQ{
y^2=(x^2-a^2-2\Lambda^2)^2-4\Lambda^4\ ,
}
which is the Seiberg-Witten curve of an $SU(2)$ theory. The potential
required is
\EQ{
W(x)=N\varepsilon\big(\frac13x^3-(a^2+2\Lambda^2)x\big)\ .
}

As $a\to0$, by hypothesis, there is a light dibaryon with
mass controlled by  
\SP{
\tilde S&=
N\int_{-a}^a\sqrt{(x^2-a^2)(x^2-a^2-2\Lambda^2)}\,dx\\
&=\frac23N
\sqrt{a^2+2\Lambda^2}\big(2(a^2+\Lambda^2)E(a^2/(a^2+2\Lambda^2))
+2\Lambda^2K(a^2/(a^2+2\Lambda^2))\big)\ .
}
At leading order for small $a$
\EQ{
\tilde S=\frac\pi{\sqrt2}N\Lambda a^2+\cdots
}
Notice that the double-scaling limit involves $N\to\infty$  with
$Na^2$ fixed, in agreement with \eqref{dsp}.
One way to ``see'' the dibaryon is to calculate the low energy 
couplings of the two $U(1)$'s. The coupling constant matrix has the form
\EQ{
\tau\MAT{1&-1\\ -1&1}
}
where $\tau$ is the period matrix of the curve which can be written 
explicitly in terms of elliptic functions as \cite{DGKV}
\EQ{
\tau=\frac{K(a^2/(a^2+2\Lambda^2))}{E(2\Lambda^2/(a^2+2\Lambda^2))}\ .
}
In the limit $a\to0$, the leading-order behaviour is
\EQ{
\tau=\frac{\pi}{\log(\alpha/a^2)}\ ,
}
for some constant $\alpha$.
This kind of logarithmic running of the coupling 
is characteristic of a one-loop effect of a 
particle of mass $\sim\tilde S$ which is magnetically charged under 
the two $U(1)$'s. 

However, in this case we can also extract the behaviour of the higher
genus $F_g$ as $a\to0$. In the present
case, near the critical point we can approximate the curve by
\EQ{
y^2\longrightarrow 4\Lambda^2(x^2-a^2)\ .
}
Now we can appeal to universality in the double-scaling limit. The
reduced curve above is precisely the one encountered in the 
matrix model with a Gaussian potential
\EQ{
W(x)= N \varepsilon \Lambda x^2\ .
}
So the free energy of our original matrix model in the double-scaling
limit should be equal to the free energy of the Gaussian model with
the 't~Hooft coupling $S=g_s\hat N$ of the latter identified with  
$\tilde S$ above. The Gaussian matrix model can be solved exactly
\cite{Dijkgraaf:2002fc} yielding
\EQ{
F_\text{Gaussian}(S)=\frac12g_s^{-2}S^2\log S-\frac1{12}\log S
-\frac1{240}g_s^2S^{-2}+\sum_{g>2}\frac{B_g}{2g(2g-2)}g_s^{2g-2}S^{2-2g}\ .
\label{Gauss1}}
Hence, by appealing to universality, we deduce that in the double-scaling limit 
of the original theory as $a\to0$
\EQ{
F_g\thicksim \tilde S^{2-2g}\varpropto|Na^2|^{2-2g}\ ,
}
which verifies in a simple example the scaling hypothesis \eqref{dsp}. In
particular, notice that the logarithm in $F_0$ is directly
attributed to a one-loop renormalization of the coupling due to the
light state.

In \cite{Bertoldi:2006gk}, the methods of \cite{Eynard,ChekhovEynard,ChekhovG=1}
are used to evaluate the genus one and two terms of the matrix model
free energy for the class of double-scaling limits considered in \cite{BD}.
The simplest case corresponds to the above conifold singularity with
two branch points colliding and no double points, $n=2, m=0$,
and the results match the genus zero, one and two terms in \eqref{Gauss1}. 
This is precisely due to the fact that the near-critical spectral curve reduces
to the curve of a Gaussian matrix model, which is simply a Riemann sphere.

\subsection{Orthogonal polynomials.}

Motivated by the study of random surfaces and 2-d quantum gravity, the double scaling limits of multi-cut matrix models were investigated  in the early 90's using orthogonal polynomials.
This technique was originally developed in \cite{L_Bessis} as an alternative way to calculate the matrix model free energy
\EQ{
\exp\left( F\right) =\int
d\hat\Phi\,\exp\Big(-g_s^{-1}\text{Tr}\,W(\hat\Phi)\Big)
=\int \prod_{i=1}^{\hat N}\> dx_i\> \Delta^2(x)\> {\rm e\>}^{-g_s^{-1} \sum_i W(x_i)}\>,
\label{L_matrix}
}
where $\Delta(x)=\prod_{i\not=j}(x_i-x_j)$ is the Vandermonde determinant, and its main features can be found in the reviews \cite{DiFrancesco:1993nw,L_Marinho}. The basic idea is to calculate the integral in~\eqref{L_matrix} by introducing the set of polynomials $P_n(x)$ orthogonal with respect to the measure 
$\int dx\; {\rm e\>}^{-g_s^{-1} W(x)}\> P_n(x) 
\> P_m(x) = h_n \delta_{mn}
$, where $P_n(x)$ is a polynomial of degree $n$ normalized such that $P_n(x)=x^n +\cdots$. Orthogonal polynomials satisfy recursion relations of the form
$x P_n(x)= P_{n+1}(x) + \sigma_n P_n(x) + r_n P_{n-1}(x)$, where $r_n$
and $\sigma_n$ are $x$-independent purely numerical coefficients, and $\sigma_n=0$ if the potential is even. They can be determined by solving the non-linear recursive equations
\begin{eqnarray}
&&
n\>g_s\> h_{n-1} = \int \> {\rm e\>}^{-g_s^{-1} W(x)}\> W'(x)\> P_{n}(x)\> P_{n-1}(x)\>, \label{L_recursion1}\\[5pt]
&&
0=\int \> {\rm e\>}^{-g_s^{-1} W(x)}\> W'(x)\> P_{n}(x)\> P_{n}(x)\label{L_recursion2}\>.
\end{eqnarray}
In the large-$\hat N$ limit, the re-scaled index $n/{\hat N}$ becomes a continuous variable $\xi\in [0,1]$, and the orthogonal polynomial method becomes useful provided that we have an appropriate ansatz for the large-$n$ behaviour of $r_n$ and $\sigma_n$. 
In the double scaling limit, the method provides important information
about the resulting models. The reason is that the free energy turns
out to be described in terms of particular solutions of specific
classical integrable hierarchies. The relevant solutions are singled
out by additional equations known as  ``string equations''. This
important result permits the calculation of the complete perturbative
(topological) expansion of the free energy in terms of the double
scaled parameters, and it also provides non-perturbative information
about the resulting models. Moreover, the relationship with integrable
hierarchies makes it possible to write flow equations that interpolate between different models.

The simplest large~$n$ ansatz is $r_n\rightarrow r(\xi;S)$ and $\sigma_n\rightarrow \sigma(\xi;S)$, such that both coefficients become smooth functions of $\xi$ and the 't~Hooft coupling $S=g_s{\hat N}$. It corresponds to saddle point configurations whose large-$\hat N$ resolvent has just one cut \cite{L_Bessis,L_TwoArcs1}. In this case, to describe all the resulting double scaling limits, it is enough to restrict ourselves to the case of Hermitian matrix models with even potentials. Then, the topological large-$\hat N$ expansion of the free energy exhibits a singular behaviour of the form
\EQ{
F= \sum_{g\geq0} F_g(S)\> {\hat N}^{2-2g}\>, \qquad F_g(S) \sim (S-S_c)^{(2+{1\over m+1})(1-g)}\>, \quad m\geq1\>,
\label{L_OneCrit}
}
up to a few regular terms in $F_0$. In this approach it is customary
to keep the potential $W$ fixed and consider just the dependence on
$S$; however, we could equivalently fix the value of $S=S_c$ and look
at the critical behaviour as a function of the coupling constants in
$W$.\footnote{This is exactly what it done in the Dijkgraaf-Vafa
  matrix model, where $S$ is fixed by the requirement to extremize the
  glueball superpotential.} Adopting the first point of view, both $S_c$ and the integer $m$ depend on $W$. In the saddle point method, the critical behaviour~\eqref{L_OneCrit} corresponds to the case when $m$ double points come together with one of the branch points of the resolvent, which can be realized with potentials of degree larger or equal $2(m+1)$. It leads to the double scaling limit
$\hat N\rightarrow \infty$,  $\>S\rightarrow S_c $, with $ {\hat N} (S-S_c)^{1+{1\over 2(m+1)}} = \text{finite}$. Using orthogonal polynomials, this limit can be performed  with a scaling ansatz of the form
\EQ{
r(\xi;S) = r_c\left(1 + {u(z)\over \; {\hat N}^{ {2\over 2m+3}}\;} + \cdots  \right)\>, \qquad \sigma(\xi;S)=0\>,
}
together with $n g_s= \xi S = S_c\left(1-{z\> {\hat N}^{- {2m+2\over 2m+3}}} \right)$ . 
It is worth noticing that, in the saddle point approximation, this ansatz corresponds to a resolvent with two branch points symmetrically located at~\cite{L_Bessis,L_TwoArcs1}
\EQ{
x=\pm x_b=\pm\sqrt{r(1;S)}\sim \pm\bigl(x_c + O({\hat N}^{- {2\over 2m+3}})\bigr)\>.
}
Then, 
\EQ{
a=x_b-x_c \sim \bigl(1-S/S_c\bigr)^{1\over m+1}={\hat N}^{-{2\over 2m+3}}\; z^{1\over m+1}\>,
}
and the double scaling limit prescription can be written as ${\hat
  N}a^{m+{3\over2}}= \text{const.}$, which is in agreement with
eq. (2.12), and ensures that the free energy scales as in~\eqref{dsp} with $n=1$.
The resulting models are well known~\cite{DiFrancesco:1993nw}. The free energy is given by $d^2 F /dz^2 =u$, where $u=u(z)$ is a particular solution of the KdV hierarchy that satisfies a string equation depending on~$m$. In the simplest case, $m=1$,
the string equation is
$z= u^2 +u''/3$,
which is known as Painlev\'e~I; the resulting model describes pure 2-d gravity. For $m>1$ the model corresponds to the $(2m+1,2)$ CFT minimal model coupled to 2-d gravity. It can be shown that the $m$-th string equation is associated to the $m+1$-th flow of the KdV hierarchy. Although in the matrix model approach $m\geq1$, the integrable hierarchy formulation allows one to construct an additional model associated to the $1^\text{st}$ flow of the hierarchy. Actually it corresponds to a topological theory that underlies all the other double scaled models, and it was identified with 2-d topological gravity~\cite{L_TopGrav}.

Saddle point configurations with two cuts are recovered with a ``period-two'' ansatz; {\em i.e.\/}, by assuming that $r_{2n}$, $r_{2n+1}$, $\sigma_{2n}$ and $\sigma_{2n+1}$ approach different smooth functions in the large-$n$ limit~\cite{L_Period2}.
However, the resulting cuts always open up around minima of the
potential $W(x)$~\cite{L_TwoArcs1,L_TwoArcs2} and it is not known how to describe saddle point configurations with cuts open around the maxima of $W(x)$ using orthogonal polynomials. This includes, for instance, the two-cut phase of the matrix model specified by $W={1\over 3}g_3x^3 +{g_1}x$, which was extensively discussed in~\cite{DGKV,KMT}, as well as in section 4.1. Actually, this limitation poses a serious restriction for the use of the orthogonal polynomial method in the context of~\eqref{L_DV}. In any case, it is interesting to review the main results achieved by considering the class of two-cut matrix models that can be studied using this method.

Two-cut Hermitian matrix models with even potentials were investigated in~\cite{L_CMoore1}. They exhibit critical behaviours corresponding to cases where two branch points come together with an odd number of double points, and lead to the same double scaling limits that were previously found in the study of unitary matrix models~\cite{L_Unitary}, which are described in terms of the modified KdV hierarchy. 
The interpretation of the resulting models in terms of super CFTs coupled to 2-d supergravity was suggested in the early 90's~\cite{L_CMoore1,L_CMoore2}, and it has been recently clarified and supported by Klebanov, Maldacena and Seiberg~\cite{L_Klebanov:2003wg}.
However, these are not the only two-cut models that have been studied
using orthogonal polynomials. Two-cut Hermitian matrix models with
both even and odd terms in the potential were considered
in~\cite{L_CMoore2,Nappi:1990bi,L_Oldpap}. Remarkably, the
corresponding double scaling limits are described in terms of several
integrable hierarchies associated to $sl(2,{\bf C})$. Following~\cite{L_Oldpap}, the results can be summarized as follows. First, consider the case of real potentials. In the saddle point approximation, their critical behaviours correspond to configurations where an odd number of double points come together with two branch points~\cite{L_CMoore2,L_Oldpap}. Their double scaling limits are described in terms of solutions of the non-linear Schr\"odinger (NLS) hierarchy singled out by string equations associated only to the ``even'' flows of the hierarchy. 

However, as pointed out in~\cite{L_Oldpap}, the most general critical behaviour, where an arbitrary number of double points collide with two branch points, is obtained by considering complex potentials. Namely, the potential has to be taken such that $W^\ast(x)=W(-x)$, which makes the coefficients $\sigma_n$ pure imaginary.\footnote{This is equivalent to consider anti-Hermitian matrix models, whose eigenvalues are purely imaginary, with real potentials.} This gives rise to double scaling limits described in terms of solutions of the Zakharov-Shabat (ZS) hierarchy specified by string equations associated to the complete set of flows, but the first one. The resulting models are described by two real functions $f(z)$ and $g(z)$. They enter the ansatz for the orthogonal polynomial coefficients, which is of the form
\EQ{
r_n = r_c\left(1 + (-1)^n{f(z)\over \;{\hat N}^{{1\over m+2}}\;} + \cdots  \right)\>, \quad
\sigma_n = i\sigma_c\left(b + (-1)^n{g(z)\over \;{\hat N}^{{1\over m+2}}\;} + \cdots  \right)\>,
}
together with $\xi S = S_c\left(1-{z\> {\hat N}^{- {m+1\over m+2}}} \right)$. Here, $b$ is an arbitrary number, and $m$ is the number of double points.
In the saddle point approximation, this ansatz gives rise to a resolvent where the two colliding branch points are located at~\cite{L_TwoArcs1,L_TwoArcs2}
\begin{eqnarray}
x=x_b^\pm&=&
{\sigma_{2n}+\sigma_{2n+1} \pm \sqrt{\bigl(\sigma_{2n}-\sigma_{2n+1}\bigr)^2 -4\bigl( \sqrt{r_{2n}}- \sqrt{r_{2n+1}}\bigr)^2}\over 2
}\;\bigg|_{\xi=1}\nonumber\\[5pt]
&\sim& ib\sigma_c \pm O({\hat N}^{-{1\over m+2}})\>.
\end{eqnarray}
Thus, in this case,
\EQ{
a= x_b^+ -x_b^- \sim \bigl(1-S/S_c\bigr)^{1\over m+1}={\hat N}^{-{1\over m+2}}\; z^{1\over m+1}\>,
}
and the double scaling limit prescription can be written as ${\hat N}a^{m+2}= \text{const.}$, which is again in agreement with eq. (2.12) for $n=2$. The free energy is given by 
$d^2 F/d^2 z = \bigl(f^2(z)-g^2(z)\bigr)/4$. For $m=1$ the string equations are
\EQ{
zf +f(g^2-f^2) + 2f''=0\>, \qquad zg +g(g^2-f^2) + 2g''=0\>,
}
and for $m=2$ they read 
\EQ{
2zf +3g'(g^2-f^2) + 2g'''=0\>, \qquad 2zg +3f'(g^2-f^2) + 2f'''=0\>.
}
They exhibit that the string equations corresponding to odd $m$ admit the reduction $g=0$ and the analytical continuation $g\rightarrow ig$, which lead to the mKdV and the NLS hierarchies, respectively. Other reductions, as well as the relationship between the partition function and the tau-functions of the hierarchies, have been discussed in~\cite{L_Oldpap}.
It is straightforward to check that the free energy scales exactly as
in \eqref{dsp} with $n=2$. 

An interesting observation about the structure of these two-cut matrix models was made in~\cite{L_CMoore2}. There, it was shown that the formulation in terms of the ZS hierarchy allows one to define a model associated to the $1^\text{st}$ flow which is of topological nature. This is similar to what happens in the one-cut case, where the $1^\text{st}$ flow of the KdV hierarchy leads to 2-d topological gravity. By comparison, it is natural to expect that the topological model of~\cite{L_CMoore2} underlies the structure of all the ZS-related two-cut matrix models, although the proper interpretation of this topological phase is still unclear.

Another important feature pointed out in~\cite{L_TwoArcs2} and, mostly, in \cite{L_Klebanov:2003wg} is the fact that the string equations of these models admit an integration constant. For the models associated to the ZS hierarchy, this is a consequence of the invariance under the $SO(1,1)$ transformations $f\pm g\rightarrow {\rm e\>}^{\pm \beta}(f\pm g)$. This constant shows up in the perturbative expansion of the free energy, although the leading order term remains independent of it. The only exception is the topological model of~\cite{L_CMoore2}, where the free energy is proportional to a non-vanishing integration constant. However, in~\cite{L_CMoore2}  the presence of this constant in the non-topological models was missed, mainly because it was assumed that the potential is real and, consequently, the colliding double points and branch points are on the real axis.
In \cite{L_Klebanov:2003wg}, double points were allowed to be complex, the integration constant was understood in terms of the period of the hyper-elliptic $y=y(x)$ curve of the matrix model around the cut joining the two colliding branch points, and its physical interpretation was clarified. When $g\rightarrow ig$, and the ZS hierarchy changes to the NLS hierarchy,
the constant becomes complex too, and it can be related to the difference between the number of eigenvalues sitting on the two cuts \cite{L_TwoArcs2,L_Klebanov:2003wg}.

Motivated by the correspondence between the period-2 large-$n$ ansatz
and two-cut saddle point configurations, a study of multi-cut
configurations by considering more general period-$q$ large-$n$
behaviours; {\em i.e.\/}, $r_{lq+i}\rightarrow r_i(\xi;S)$ and
$\sigma_{lq+i}\rightarrow \sigma_i(\xi;S)$, for $i=0,1,\ldots,q-1$,
was initiated. However, for $q\geq3$ this sort of ansatz is only able to reproduce multi-cut saddle point configurations for very special types of potentials. For instance, for $q=3$ and even potentials of degree six with three minima, the correspondence turns out to works only in the particular case when all the minima are degenerate~\cite{L_TwoArcs1}. In other words, the appropriate large-$n$ ansatz corresponding to generic saddle point configurations with three or more cuts in not known, which prevents the use of the orthogonal polynomial method to investigate the properties of these configurations and their double scaling limits. Several aspects of this problem were discussed in~\cite{L_Chaos}.

An interesting particular set of multi-cut matrix models partially studied using orthogonal polynomials are provided by the, so-called, ``orbifold (Hermitian) matrix models'' proposed in~\cite{L_CMoore1}.
They give rise to saddle-point configurations with $2k$ cuts symmetrically radiating from the origin. 
The critical behaviours considered in that paper correspond to cases
where the end points of all the cuts simultaneously collide together
with $2km-1$ double points, for $m\geq1$. The resulting double-scale
models are expected to exhibit a ${\bf Z}_k$ symmetry.
The simplest potential that gives rise to such configuration is
$W(x)=x^8-x^4$ for a matrix model defined over matrices $\hat\Phi$
such that ${\hat\Phi}^2$ is hermitian, for which there are four branch
points and three double points in the critical region.
This model was studied in~\cite{L_CMoore1} by means of a very
symmetrical period-4 large-$n$ ansatz for the coefficients
$r_n$. However, both its physical interpretation and the associated
integrable hierarchy remain mysterious still. Nevertheless,  it can
verified from the resulting generalized string equations, 
that the free energy scales as in \eqref{dsp} with $n=2k$ and
$m=2km-1$. 

Apart from the aforementioned ${\bf Z}_k$ models, the fact that the
orthogonal polynomial method seems to work for one or two branch cuts
may be related to the fact that resulting near-critical curves are both
of genus zero. In fact, this is related to the fact that the string
equations in both cases can be formulated in terms of the Virasoro
algebra of a chiral boson
on the complex plane. For the case of one branch point there is a
branch cut extending to infinity and the boson is twisted,
$\phi(ze^{2\pi i})=-\phi(z)$.  

\subsection{The loop equations}

In this section, consider a radically different approach for
investigating the free energy which involves a recursion relation for
the $F_g$. Before we come to the loop equation itself we need to
define the $p$-loop correlator, or $p$-point loop function, as
\SP{
W(x_1,\ldots,x_p) 
&\equiv {\hat N}^{p-2} \Big\langle \, \text{Tr} \frac{1}{x_1-\hat\Phi}
\cdots \text{Tr} \frac{1}{x_p - \hat\Phi} \, \Big\rangle_\text{conn}\\
&= \frac{d}{dV}(x_p)  \cdots
\frac{d}{dV}(x_1) F \ , \qquad  p\geq 2\ .
}
It has the following genus expansion
\EQ{
W(x_1,\ldots,x_p) = \sum_{g=0}^\infty \frac{1}{\hat N^{2g}}
W^{(g)}(x_1,\ldots,x_p) \ .
}
In \cite{Eynard}, Eynard found a solution to the matrix model loop
equations that allows to write down an expression for these
multi-loop correlators at any given genus in terms of a special set of
Feynman diagrams.    
The various quantities involved depend only on the spectral curve of the 
matrix model and in particular one needs to evaluate residues of certain
differentials at the branch points of the spectral curve.  

This algorithm and its extension to calculate 
higher genus terms of the matrix model free energy \cite{ChekhovEynard}
represent major progress in the solution
of the matrix model via loop equations 
\cite{Makeenko,ACKM,Akemann,AmbjornAkemann,KMT,DST,ChekhovG=1}. 
This is particularly important because,as we have seen in the last section, 
the orthogonal polynomial approach can only applied in very special cases.
A nice feature of the loop equation approach is that they show directly 
how the information is encoded in the spectral curve.
In particular, we will be able to make some precise statements on 
the double-scaling limits of higher genus quantities simply
by studying the double-scaling limit of the spectral curve and its
various differentials.  

Given the matrix model spectral curve for an $s$-cut solution in the
form \eqref{mc} the genus zero $2$-loop function is given by
\SP{
W(x_1,x_2) 
&= -\frac{1}{2(x_1-x_2)^2} +
\frac{\sqrt{\sigma(x_1)}}{2\sqrt{\sigma(x_2)}(x_1-x_2)^2}\\
&-\frac{\sigma'(x_1)}{4(x_1-x_2)\sqrt{\sigma(x_1)}\sqrt{\sigma(x_2)}}
+ \frac{A(x_1,x_2)}{4\sqrt{\sigma(x_1)}\sqrt{\sigma(x_2)}}\ ,  
\label{2loop}
}
where $A$ is a symmetric polynomial given by 
\EQ{
A(x_1,x_2) = \sum_{i=1}^{2s} \frac{ {\cal L}_i(x_2) \sigma(x_1)
}{x_1-\sigma_i}\ , 
\label{A12}
}
with
\EQ{
{\cal L}_i(x_2) = \sum_{l=0}^{s-2} {\cal L}_{i,l}x_2^l =  -
\sum_{j=1}^{s-1} L_j(x_2) \int_{A_j} \frac{dx}{\sqrt{\sigma(x)}}
\frac{1}{(x-\sigma_i)}
\label{LLx}
}
and $s$ is the number of cuts. The polynomials 
$L_j(x)$ are related to the holomorphic 1-forms and defined in the Appendix.

The genus zero $2$-loop function for coincident arguments is
\SP{
W(x_1,x_1)& = \lim_{x_2 \to x_1} W(x_1,x_2) = -\frac{\sigma''(x_1)}{8
\sigma(x_1)} + \frac{\sigma'(x_1)^2}{16 \sigma(x_1)^2} +
\frac{A(x_1,x_1)}{4 \sigma(x_1) }\\
&= \sum_{i=1}^{2s} \frac{1}{16 (x-\sigma_i)^2}
- \frac{\sigma_i''}{16 \sigma'_i(x-\sigma_i)}
+ \frac{{\cal L}_i(x)}{4(x-\sigma_i)}\ .
\label{2loopx1x1}
}
The other important object is the differential
\SP{
&dS_{2i-1}(x_1,x_2) = 
dS_{2i}(x_1,x_2)\\
&= \frac{\sqrt{\sigma(x_2)}}{\sqrt{\sigma(x_1)}} \left(
\frac{1}{x_1-x_2} - \frac{L_i(x_1)}{\sqrt{\sigma(x_2)}} -
\sum_{j=1}^{s-1} C_j(x_2) L_j(x_1) \right) dx_1\ ,
\label{dSi}
}
where $i=1,\ldots, s$ and 
\begin{equation}
C_j(x_2) = \int_{A_j} \frac{dx}{\sqrt{\sigma(x)} }
\frac{1}{(x-x_2)}\ .
\label{Cj}\end{equation}
A crucial aspect of the one-form (\ref{dSi}) is that it is analytic in 
$x_2$ in the limit $x_2 \to \sigma_{2i-1}$ or $\sigma_{2i}$ \cite{Eynard}
\begin{equation}
\lim_{x_2 \to \sigma_i} \frac{dS_i(x_1,x_2)}{\sqrt{\sigma(x_2)}} =
\frac{1}{\sqrt{\sigma(x_1)}} \left( \frac{1}{x_1-x_2} -
\sum_{j=1}^{s-1}  L_j(x_1) \int_{A_j} \frac{dx}{\sqrt{\sigma(x)}}
\frac{1}{(x-x_2)} \,\, \right) dx_1\ .
\label{dSilimit}\end{equation}
The subtlety is that in the definition of (\ref{Cj}),
the point $x_2$ is taken to be outside the loop surrounding the $j$-th cut,
whereas in (\ref{dSilimit}), $x_2$ is inside the contour.
Note also that
\begin{equation}
A(x_1,x_2) = - \sum_{i=1}^{2s} \left( \sum_{j=1}^{s-1} L_j(x_2)
C_j(\sigma_i) \right) \frac{\sigma(x_1)}{x_1 - \sigma_i}
\label{A12explicit}\end{equation} 
and in particular
\begin{equation}
A(x_1,\sigma_i) = {\cal L}_i(x_1) \sigma'(\sigma_i) \ .
\label{A12explicit2}\end{equation}

The expression of $W^{(g)}(x_1,\ldots,x_p)$ can be found
by evaluating a series of Feynman diagrams of a cubic field 
theory on the spectral curve \cite{Eynard}. To this end,
define the set ${\cal T}^{(g)}_p$ of all
possible graphs with $n$ external legs and with $g$ loops.
They can be described as follows: draw all {\it rooted\/} skeleton 
trees ( trees whose vertices have valence $1,2$ or $3$ ),
with $p+2g-2$ edges. Draw arrows on the edges
from the root towards the leaves. Then draw in all possible ways
$p-1$ external legs and $g$ inner edges with the constraint that
all the vertices of the whole graph have valence three, namely
that are always three and only three edges emanating from any given 
vertex. Each such graph will also have some symmetry factor \cite{Eynard}.

Each diagram in then weighted in the following way.
To each arrowed edge that is part of the skeleton tree going from a vertex
labelled by $x_1$ to a vertex labelled by $x_2$ associate the differential 
$dS(x_1,x_2)$ \eqref{dSi}. 
To each non-arrowed edge associate a genus zero $2$-loop
differential $G(x_1,x_2)=W(x_1,x_2)\,dx_1\,dx_2$
and to each {\it internal\/} vertex labelled by $x_1$ associate the 
factor $(2\varepsilon Ny(x_1) dx_1)^{-1}$.  
For any given tree $T \in {\cal T}_p^{(g)}$, with root $x_1$ 
and leaves $x_j, j=2,\ldots,p$ and with $p+2g-2$ vertices 
labelled by $x'_v, v=1,\ldots,p+2g-2$, so that its inner edges are of the
form $v_1 \to v_2$ and its outer edges are of the form $v \to j$,
we define the weight of the graph as follows
\SP{
{\cal W}(T) &= \frac1{(\varepsilon N)^{p+2g-2}}
\prod_{v=1}^{p+2g-2} \sum_{i_v=1}^{2s} \text{Res}_{x'_v \to b_{i_v}}\,
\frac{1}{2 y(x'_v)dx'_v} 
\prod_{\text{inner edges } v \to w} dS_{i_v}(x'_v,x'_w) \\
&\times\prod_{\text{inner non-arrowed edges }v' \to w'} G_2(x'_{v'},x'_{w'})
\prod_{\text{outer edges }v \to j} G_2(x'_v,x_j)   
\label{weight}
}
In order to find an expression for $F_g$, $g>1$, 
one should consider the same graphs relevant for
$W^{(g)}(x_1)$ and do then the following \cite{ChekhovEynard}:

\noindent
{\bf (i)} Eliminate the first arrowed edge of the skeleton tree.
Labelling the first vertex by $x_1$ and the second vertex by $x_2$,
this amounts to dropping the factor $dS(x_1,x_2)$. \\ 

\noindent
{\bf (ii)} The factor $(2\varepsilon Ny(x_2) dx_2)^{-1}$ has to be dropped 
and  replaced by 
\EQ{
\frac{\int^{x_2}_{q_0} y(s)ds}{y(x_2)dx_2}\ .
}
Note that when evaluating the final residues at $x_2 = \sigma_i$, one needs to expand
the above integral by setting $q_0 = \sigma_i$ \cite{ChekhovEynard}.  
It is also understood that the evaluation of the residues starts from the outer
branches and proceeds towards the root. 
This procedure does not apply for the genus one free energy whose expression
has in any case been found via the loop equations in \cite{KMT,DST,ChekhovG=1}.

We will consider the $a\to0$ limit of each element in \eqref{weight}.  
Using the results in the Appendix for the scaling of $L$, it is
straightforward to argue that for a branch point $b_i$ in the critical
region 
\SP{
&dS_i(x_1,x_2)\longrightarrow
d\tilde{S}_i(\tilde x_1, \tilde x_2)\\ 
&=\frac{\sqrt{\tilde{B}(\tilde{x}_2)}}{\sqrt{\tilde{B}(\tilde{x}_1)}} 
\left(
\frac{1}{\tilde{x}_1-\tilde{x}_2} 
- \frac{\tilde{L}_i(\tilde{x}_1)}{\sqrt{\tilde{B}(\tilde{x}_2)}} -
\sum_{j=1}^{p} 
\tilde{C}_j(\tilde{x}_2) \tilde{L}_j(\tilde{x}_1) \right) d
\tilde{x}_1\ ,
\label{dSiDSL}
}
where $d\tilde{S}_i$ is the analogous differential 
on $\Sigma_-$ and $L_j(x)\to a^{n/2-1}\tilde L_j(\tilde x)$ for
$j\leq[n/2]$. 
Conversely, the differentials $dS_i(x_1,x_2)$ where $i$ labels a branch point
of the spectral curve that remains outside of the critical region
give a vanishing contribution in the double-scaling limit. 
Likewise using equations \eqref{2loop}, \eqref{A12}, \eqref{LLx} and 
\eqref{A12explicit} we have
\EQ{
G(x_1,x_2) = W(x_1,x_2) \,dx_1 dx_2 \quad 
\longrightarrow 
\quad \tilde{G}(\tilde{x}_1,\tilde{x}_2) 
= \tilde{W}(\tilde{x}_1, \tilde{x}_2) \,d\tilde{x}_1 d\tilde{x}_2\ ,
}
where $\tilde{G}(\tilde{x}_1,\tilde{x}_2)$ is exactly the $2$-point 
loop correlator on $\Sigma_-$. 

So far we have seen that the double points of the 
near-critical curve do not play a role
in taking the limit of the differentials.
However, this is not the case for the final two elements of the Feynman rules 
\EQ{
y\, dx \longrightarrow\sqrt{C} 
a^{m+n/2+1}\,\, y_-\, d \tilde{x}  
\label{ydxDSL}
}
and
\begin{equation}
\frac{\int_q^x y(s)ds}{y(x)dx} \longrightarrow 
\frac{\int_{\tilde{q}}^{\tilde{x}} 
y_-(\tilde{s})d\tilde{s}}{y_-(\tilde{x})d\tilde{x}} \ .
\label{intyovery}\end{equation}
To summarize: what we have found is that all the relevant quantities
reduce to the analogous quantities on the near-critical curve in the
limit $a\to0$. In particular, being careful with the overall scaling,
the genus $g$ free energy has the limit
\EQ{
F_g\longrightarrow C^{1-g}
\Delta^{2-2g}\, F_g(\Sigma_-) \ . 
\label{FgDSLsigmam}
}
where we have emphasized that $F_g(\Sigma_-)$ 
depends only on $\Sigma_-$. This
is the result advertised in \eqref{dsp} and the property of universality.
Similarly, the genus $g$ $p$-point loop functions have the limit
\begin{equation}
W_g(x_1,\ldots,x_p) \,dx_1 \cdots dx_p 
\longrightarrow 
C^{1-g-p/2}
\Delta^{2-2g-p} \,\tilde{W}_g(\tilde{x}_1,\ldots,\tilde{x}_p) 
\,d \tilde{x}_1 \cdots d \tilde{x}_p \ .
\label{WgDSL}\end{equation}

\section{The Physics of the Double Points}\label{PhysicsDoublePoints}

In this section, we consider in more detail the case when the
near-critical curve has double points, since as we have already argued
the physics is very different from that considered in \cite{BD}
with only branch points.

It will turn out that the existence of the double points has a
profound effect on the physics of the critical
region. In the case of branch points, dibaryons
become massless when the branch points collide. 
We want to argue that certain states also become massless 
when a double point collides with a branch point. In particular the
mass scale when $m$ double points and $n$ branch points come together is
$Na^{m+n/2+1}$. In order to tease out the physics of these light states
let us consider the case with only a single branch point and one
double point. The curve in the near critical region has the form
\EQ{
y_-^2=(\tilde x-\tilde z)^2(\tilde x-\tilde b)\ .
}
This is precisely the kind of critical curve that appeared in the old
matrix literature \cite{DiFrancesco:1993nw}. 

\begin{figure}[ht]
\centerline{\includegraphics[width=3in]{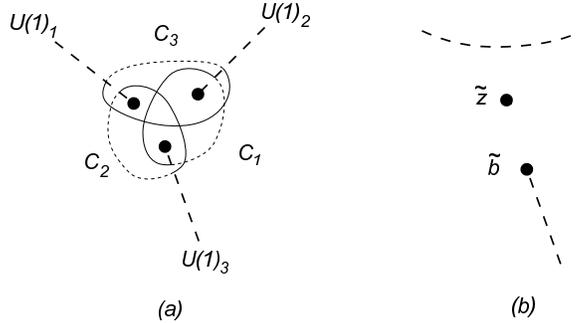}}
\caption{\footnotesize (a) A configuration for which $\Sigma_-$
  consists of three branch points. (b) Two of the branch points
  coalesce to form $\Sigma_-$ with a single branch point $\tilde b$ 
and a single double point $\tilde z$.}
\label{three}
\end{figure}

In order to motivate the issues involved, let us
consider what happens as we go through a transition where the 
double point arises from the collision of two branch
points. This can be achieved by using a more general form of the
bare superpotential than the ones we considered in Sections 2.2 and 2.3. 
In this case, on one side of the transition
we have three branch cuts ending in the critical region
and three light dibaryons associated to the three cycles $C_i$ which link
each pair of branch points. Each of the cuts is associated to an
abelian gauge field $U(1)_i$
in the IR gauge group. The situation is illustrated in Figure \ref{three}.
The dibaryon $C_i$ is magnetically,
and potentially electrically,
charged with respect to $U(1)_j\times U(1)_k$. In fact,
denoting the electric and magnetic charges as
$(g_1,q_1;g_2,q_2;g_3,q_3)$ we can choose conventions where the 
three dibaryons have charges
\SP{
C_1&:\qquad (0,0;1,1;-1,0)\\
C_2&:\qquad (-1,0;0,-1;1,0)\\
C_3&:\qquad (1,0;-1,0;0,0)\ .
}
The three dibaryons have mutually non-local charges since as cycles
$C_1\cdot C_2 = C_2 \cdot C_3 =  C_3 \cdot C_1=1$ 
and $C_1+C_2+C_3=0$ \cite{AD}. Now suppose two
branch points come together corresponding to the vanishing of
$C_3$. At this point the dibaryon $C_3$ becomes
massless and the double points of the $\N=1$ and
Seiberg-Witten curves coincide. This point in the 
parameter space touches a new phase where the dibaryon is 
condensed---the double points of the two curves move apart---and 
the $U(1)$ subgroup of $U(1)_1\times
U(1)_2$ defined by the elements $(e^{i\theta},e^{-i\theta})$ is
confined. This new confined phase is
the one we are interested in. Since dibaryons $C_1$
and $C_2$ carry opposite electric charge under the confined $U(1)$ it
is natural that $C_1$ and $C_2$ form a bound state: a ``meson''. The
bound state has minus the charges of $C_3$; in other words, the meson
is magnetically charged under the confined $U(1)$. Notice,
however, that the state is neutral under the two remaining
$U(1)$'s. This means that, in stark contrast to the case of dibaryons, 
we will not ``see'' these bound states via the
logarithmic running of the coupling of the IR abelian 
gauge fields as we did for the case of a light dibaryon in Section
\ref{DoubleFg}.  
However, as we have seen, their effects will show up in other
quantities. In particular, if this meson, of mass $M$, 
runs around 1-loop graphs
then it contributes as $M^{2-2g}$ to the higher-genus free energy
$F_g$. So the fact that $F_g\sim\Delta^{2-2g}$ means that since $m=1$
and $n=1$ the mass of
the meson is $M\sim\Delta=Na^{5/2}$, in this case.

The remaining issue is to argue that the meson becomes massless
when the double point coincides with the remaining branch point, since
as $a\to0$ evidently $M\to0$. One
might have thought that the meson has a mass of order the confinement
scale and would not become massless at the critical point. The
issue is rather reminiscent of the pion in QCD. The pion mass is much
lower than the scale of confinement because the pion is the would-be
Goldstone boson of chiral symmetry breaking. However, chiral symmetry
is explicitly broken by the mass of the light quarks and so a mass
for the pion is generated but at a scale much smaller than the
confinement scale. In the present situation, when the
double point meets the branch point {\it and\/} the Seiberg-Witten 
double point, $z=b=h$, 
the dibaryon condensate vanishes and the theory is at
an $\N=2$ Argyres-Douglas superconformal field theory \cite{AD}.
As the Seiberg-Witten double point moves away
(but leaving the branch point and double point of the $\N=1$ curve
coincident $h\neq z=b$) the dibaryon condensate develops and $\N=2$ 
superconformal symmetry is spontaneously broken. As the Seiberg-Witten
double point moves out to infinity, only an $\N=1$ superconformal
symmetry remains. Our proposal is that 
the mesonic bound-state is the
Goldstone mode of this broken symmetry and hence is exactly massless when the
branch and double points coincide. As the
double point of the $\N=1$ curve moves away from the branch point,
$z\neq b$, this
state becomes massive with a mass $\sim\Delta$ as indicated by the
behaviour of $F_g$ in the double-scaling limit \eqref{FgDSLsigmam}.

In the general case, there will be bound states of the type described 
associated to each pair consisting of a branch point and a double
point. These bound states will be magnetically charged with respect to
the confined $U(1)$ associated to each double point. In the limit we
are considering with $h_i$ finite as $a\to0$,  the confinement scale is 
always much greater than the masses of the bound states. 

\section{Conclusion}

In this paper, we have studied the large-$N$ limit of certain ${\cal N}=1$
theories in the proximity of Argyres-Douglas-type singularities.
The exact analysis of the $F$-terms performed via the 
Dijkgraaf-Vafa matrix model correspondence shows that, close to these 
singular points, the $1/N$ expansion breaks down. This can be traced back 
to the appearance of extra light particles of baryonic and/or mesonic nature
that become massless at the critical point. 
As in \cite{BD}, there is a natural large-$N$ {\it double-scaling\/} limit
that emerges: namely if one considers approaching the singularity  
in conjunction with the 't Hooft large-$N$ limit in such a way that 
the mass $M$
of these baryonic or mesonic 
states is kept fixed. In this limit, we have argued that 
the Hilbert space of the theory
splits into two mutually decoupled sectors, ${\cal H}_+$ and ${\cal H}_-$.
The sector ${\cal H}_+$ obeys the usual large-$N$ scaling and becomes
free in the limit, whilst 
the sector ${\cal H}_-$ keeps non-trivial interactions
weighted by the effective string coupling $g_\text{eff} \sim 1/M$. 
In \cite{BD}, it was possible to map this large-$N$ double-scaling limit 
to an analogous double-scaling limit considered in \cite{GK} in the context
of the duality between four-dimensional 
Little String Theories and certain non-critical string
backgrounds \cite{GKP}.  This led to the proposal that the non-trivial dynamics
of the ${\cal H}_-$  sector admits a dual holographic description in terms 
of the four-dimensional non-critical string.  
This proposal passes a non-trivial consistency check in that the
non-critical string dual exhibits ${\cal N}=2$ supersymmetry while the
$F$-term effective action was shown to be consistent with an 
enhancement from ${\cal N}=1$ to ${\cal N}=2$
supersymmetry in the ${\cal H}_-$ sector.
In this paper, we have shown that for more general singularities
the sector ${\cal H}_-$ has only ${\cal N}=1$ supersymmetry. In particular, 
the effects of the glueball superpotential do not vanish in the limit. 
Thus, even though the field theory analysis clearly shows 
that the double-scaling limit still 
selects a particular sector of the theory which
has non-trivial dynamics, we do not identify the would-be non-critical
string dual as in \cite{BD}. For the type of singularities studied 
in the old matrix model to describe two-dimensional gravity coupled to
$c < 1$ conformal field theories, where the near-critical curve 
has only one branch point along with any arbitrary number of double points, 
the double-scaling limit describes a sector of a gauge theory 
with a mass gap and light meson-like composite states which we argued are
the approximate Goldstone bosons of superconformal invariance.
These cases are special in that there are no massless abelian fields
in the ${\cal H}_-$ sector; in fact, there is a mass gap.

{\bf Acknowledgments}. We would like to thank Nick Dorey, Prem Kumar
and Asad Naqvi for discussions. 
JLM was partly supported by MCyT (Spain) and FEDER (FPA2005-00188 and           
FPA2005-01963), Incentivos from Xunta de Galicia, and the EC network EUCLID    
under contract HPRN-CT-2002-00325.

\startappendix

\Appendix{Some Formulae}

In this appendix, we consider the double-scaling limit of various
quantities defined on the curve $\Sigma$ \eqref{mc}. This is most
conveniently done in the basis $\{\tilde A_i,\tilde B_i\}$ of 1-cycles
described in Section \ref{DoubleFterms}. In particular, for $i\leq[n/2]$ these are
cycles on the near-critical curve $\Sigma_-$ in the double-scaling
limit.   

The key quantities that we will need are the periods
\EQ{
M_{ij} = \oint_{\tilde B_j} 
\frac{x^{i-1}}{\sqrt{\sigma(x)}}\, dx 
\ ,
\qquad
N_{ij} =
\oint_{\tilde A_j} \frac{x^{i-1}}{\sqrt{\sigma(x)}}\, dx \ .
\label{MijNijapp}
}
First of all, let us focus on $N_{ij}$ where $j\leq[n/2]$, but $i$ arbitrary.
By a simple scaling argument, as $a\to0$,
\EQ{
N_{ij} 
=
\int_{b_{(j)}^-}^{b_{(j)}^+} \frac{x^{i-1}}{\sqrt{B(x)}}\, dx
\longrightarrow
a^{i-n/2}
\int_{\tilde b_{(j)}^-}^{\tilde b_{(j)}^+} 
\frac{\tilde x^{i-1}}{\sqrt{ \tilde B(\tilde{x}) }}\, d \tilde x  
= a^{i-n/2}
\, f_{ij}^{(N)} ( \tilde{b}_l) \ ,
\label{N--}
}
for some function $f_{ij}^{(N)}$ of the branch points of $\Sigma_-$.
Here, $b_{(j)}^\pm$ are the two branch points enclosed by the cycle
$\tilde A_j$. A similar argument shows that $M_{ij}$ scales  in the
same way:
\EQ{
M_{ij}\longrightarrow a^{i-n/2}
\, f_{ij}^{(M)} ( \tilde{b}_l) \ .
}
So both
$N_{ij}$ and $M_{ij}$, for $i,j,\leq[n/2]$, diverge in the limit 
$a \to 0$. On the contrary, by using a similar argument, 
it is not difficult to see that, for $j>[n/2]$, $N_{ij}$ and $M_{ij}$ 
are analytic as $a\to0$ since the integrals are over non-vanishing
cycles. 

In summary, in the limit $a \to 0$, the matrices
$N$ and $M$ will have the following block structure
\EQ{
N \longrightarrow \left(
\begin{array}{cc}
N_{--} & N_{-+}^{(0)} \\
0 & N^{(0)}_{++} \\
\end{array} 
\right)\ , 
\qquad
M \longrightarrow \left(
\begin{array}{cc}
M_{--} & M_{-+}^{(0)} \\
0 & M^{(0)}_{++} \\
\end{array} 
\right) \ ,
\label{lim}
}
where by $-$ or $+$ we denote indices in the 
ranges $\{1, \ldots, [n/2]\}$ and $\{[n/2]+1,\ldots,s-1\}$
respectively. In \eqref{lim}, $N_{--}$ and $M_{--}$ are divergent
while the remaining quantities are finite as $a\to0$.

We also need the inverse $L=N^{-1}$. 
In the text, we use the polynomials $L_j(x)=\sum_{k=1}^{s-1}L_{jk}x^{k-1}$,
which enter the expression of the 
holomorphic 1-forms associated to our basis of 1-cycles,
\EQ{
\oint_{\tilde A_i}\omega_j=\delta_{ij}\  .
}
These 1-forms are equal to
\EQ{
\omega_j(x) = \frac{L_j(x)}{\sqrt{\sigma(x)}}\, dx \ =  
\frac{\sum_{k=1}^{s-1} L_{jk} x^{k-1}}{\sqrt{\sigma(x)}}\, dx \ , \qquad
\oint_{A_i} \omega_j(x) = \delta_{ij}
\label{hold}
}
where $i,j=1,\ldots,s-1$. From the behaviour of $N$ in
the limit $a\to0$, we have
\begin{equation}
L = N^{-1}
\longrightarrow \left(
\begin{array}{cc}
N^{-1}_{--} & {\cal N} \\
0 & \left( N^{(0)}_{++} \right)^{-1} \\
\end{array} 
\right) \ ,
\qquad 
{\cal N} = - N_{--}^{-1} \, N_{-+}^{(0)} \, 
\left( N_{++}^{(0)} \right)^{-1} \ . 
\label{NinvLimit}\end{equation}
Since $N_{--}$ is singular we see that $L$ is block diagonal in the
limit $a\to0$. This is just an expression of the fact that the curve
factorizes $\Sigma\to\Sigma_-\cup\Sigma_+$ as $a\to0$. In this
limit, using the scaling of elements of $L_{jk}$, we
find, for $j\leq[n/2]$,
\EQ{
\omega_j\longrightarrow \frac{\sum_{k=1}^{[n/2]}(f^{(N)})^{-1}_{jk}\tilde
  x^{k-1}}{\sqrt{\tilde B(\tilde x)}}d\tilde x=\tilde\omega_j\ .
}
the holomorphic 1-forms of $\Sigma_-$. While for $j>[n/2]$, 
\EQ{
\omega_j\longrightarrow \frac{\sum_{k>[n/2]}^{s-1}(N_{++}^{(0)})^{-1}_{jk}
x^{k-n/2-1}}{\sqrt{F(x)}}dx\ ,
}
are the holomorphic 1-forms of $\Sigma_+$.

\end{document}